\documentclass[conference]{IEEEtran}
\usepackage{lipsum}
\usepackage{graphicx}
\usepackage{subfigure}
\usepackage{amsmath}
\usepackage[export]{adjustbox}
\makeatletter
\usepackage[table,xcdraw]{xcolor}
\usepackage{graphicx,xparse,booktabs}
\makeatother
\begin{document}
\title{Visualization for Histopathology Images using Graph Convolutional Neural Networks}
\author{\IEEEauthorblockN{Mookund Sureka, Abhijeet Patil, Deepak Anand, Amit Sethi}
\IEEEauthorblockA{Indian Institute of Technology, Bombay}}
\maketitle 
\begin{abstract}
With the increase in the use of deep learning for computer-aided diagnosis in medical images, the criticism of the black-box nature of the deep learning models is also on the rise. The medical community needs interpretable models for both due diligence and advancing the understanding of disease and treatment mechanisms. In histology, in particular, while there is rich detail available at the cellular level and that of spatial relationships between cells, it is difficult to modify convolutional neural networks to point out the relevant visual features. We adopt an approach to model histology tissue as a graph of nuclei and develop a graph convolutional network framework based on attention mechanism and node occlusion for disease diagnosis. The proposed method highlights the relative contribution of each cell nucleus in the whole-slide image. Our visualization of such networks trained to distinguish between invasive and in-situ breast cancers, and Gleason 3 and 4 prostate cancers generate interpretable visual maps that correspond well with our understanding of the structures that are important to experts for their diagnosis.
\end{abstract}

\section{Introduction}
Just like with natural images, deep convolutional neural networks (CNNs) have shown impressive results for the classification of various diseases in medical images \cite{rajpurkar2017chexnet}, \cite{10.3389/fmed.2019.00264}, \cite{campanella2019clinical}. CNNs have also been used on histopathology images for tasks such as screening pre-cancerous lesions and localizing tumors \cite{spanhol2016breast}, as well as predicting mutations \cite{coudray2018classification}, survival \cite{zhu2017wsisa}, and cancer recurrence \cite{xu2019deep}\cite{ye2018hybrid}\cite{mkl}.

Though CNN based algorithms on histopathology images have produced promising results, these algorithms lack interpretability. Localization and visualization algorithms in CNNs such as guided-backpropagation \cite{springenberg2014striving}, grad-CAM \cite{selvaraju2017grad}, and other CAM-related techniques fail to produce informative visualization for histopathology images. For instance, these techniques do not highlight cell nuclei responsible for the diagnosis and relevant features of the tumor microenvironment to further our understanding of disease and treatment mechanisms. Also, often CNNs are not able to highlight the relevant portions of the macro environment of the tumor due to large sizes (giga-pixels) of the whole-slide images. 

Morphological features of nuclei and the spatial relationships between them decide the diagnosis of histopathology slide. Representing histopathology images in the form of graphs can help capture the interaction between nuclei and the spatial arrangement of the relative positions with each other. Nuclei are represented as nodes of a graph and the distance between the nuclei can be described as edges between nodes of a graph\cite{gadiya2019histographs}. This representation of histopathology images as graphs can be fed to graph convolutional networks (GCNs) to learn the characteristics of tissue at the macro-environment level.

Taking the idea of using GCNs on graphs extracted from histology images further in this work, we propose to use an attention-based architecture and an occlusion-based visualization technique to highlight informative nuclei and inter-nuclear relationships. Our visualization results for classification of disease states in breast and prostate cancer datasets agree satisfactorily with the pathologists' observations of the relevance of various inter-nuclear relationships. Our technique paves the way for visualization of previously unknown features relevant for more important problems such as prognosis and prediction of treatment response.

\section{Related Work}
Before the emergence of deep learning, processing of histopathology images as graphs was explored in various ways. Weyn et al.\cite{weyn1999computer} represents a histopathology image as a minimum spanning tree for the diagnosis of mesotheliomas. They use k-nearest neighbor for the classification of minimum spanning trees. Similarly, Cigdem et al. \cite{demir2005augmented} form a graph from a histopathology image by considering the cluster of nuclei as a node that is connected using binary edges between nodes. A multi-layer perceptron is used for the detection of inflammation in brain biopsy. Cell-graphs \cite{yener2016cell} uses nuclei as nodes and heuristic features as node and vertex features to perform classification on breast cancer and brain biopsy datasets.

\begin{figure*}
\centering    
\includegraphics[width=0.9\linewidth]{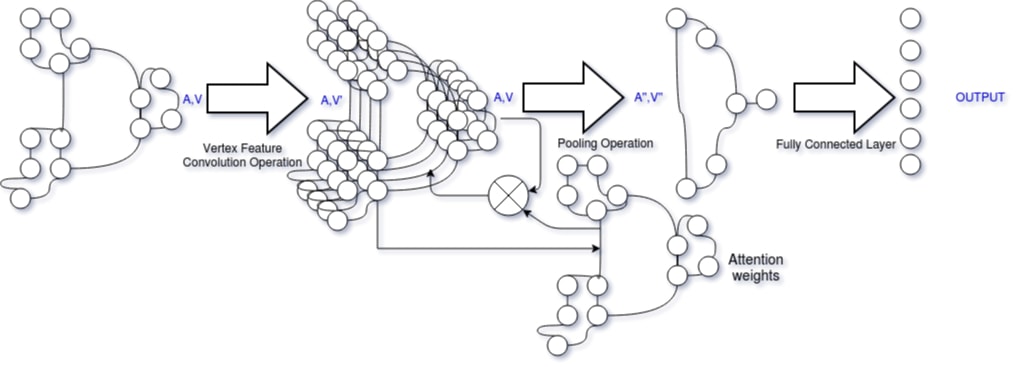}
\caption{Proposed graph convolutional neural network with an attention layer.}
\end{figure*}
Though the above mentioned methods form graphs from histopathology images, they use classical machine learning approaches such as support vector machine (SVM),k-nearest neighbors (kNN), etc. Recent developments in deep learning for graphs have enabled the use of GCNs on graphs derived from histopathology images. Kipf et al. \cite{kipf2016semi} exhibits impressive results for node classification on various graph datasets such as Citeseer, Cora, Pubmed and NELL. They used spectral graph convolution to operate on homogeneous graphs. Other lines of work in GCNs operate in the spectral domain, which enables these algorithms to analyze heterogeneous graphs as well. Such et al. \cite{such2017robust} introduced a graph convolutional algorithm in spatial domain. This method achieves excellent performance on various graph datasets. CGC-Net\cite{zhou2019cgc} uses a variant of GraphSage\cite{hamilton2017inductive} for identification of grade of prostate cancer slide represented as a graph. Recently, GCNs have been applied to graphs of nuclei in histopathology images with classification accuracy that is at par with CNNs \cite{gadiya2019histographs}.

A large portion of the medical community is skeptical about deep learning deployment in histopathology due to the lack of transparency in its working. Some attempts have been made to make deep learning more explainable. For instance, attention-based multiple instance learning \cite{ilse2018attention} frames classification of histopathology images as weakly supervised problem and assigns weights to patches of a large image. This method produces an attention map for histopathology images to highlight patches important for the classification of the overall slide, but it cannot be scaled to giga-pixel images because of its substantial computation requirements. Visualization in the form of clustering and heatmaps was presented in \cite{coudray2018classification}, but insightful interpretations beyond the highlighting of the tumor regions cannot be derived through these visualizations. Not only does interpretable visualization in general for histopathology images remains an open problem, to our knowledge, visualization for histopathology images through graph representation has also not been explored yet.

\section{Datasets and Methodology}
In this section, we describe the datasets and methodology used.

\subsection{Datasets}
In order to test the ability of the proposed method to highlight interpretable features automatically, we used two datasets for which we knew the features that were expected to be seen by the pathologists. The first dataset is from ICIAR2018 Grand Challenge on Breast Cancer Histology images (BACH) \cite{aresta2019bach} and it comprises of 400 histopathology images of breast cancer. Each image of this dataset is of the size of 2048 x 1536 pixels. The original BACH dataset contains four classes, viz. normal, benign, in-situ and invasive. We trained a GCN to perform the binary classification task between invasive and in-situ classes because these two differ in the spatial arrangement of nuclei even though the nuclei themselves share similar morphologies. We used PyTorch package for our simulations.

Gleason grade classification and visualization tasks were also performed on a prostate cancer dataset~\cite{arvaniti2018automated}. This dataset consists of a total of 1506 images for various prostate cancer tumor grades. Experiments were carried out for binary classification between Gleason grade 3+3 (primary+secondary) versus Gleason grade 4+4 or 4+5.

\subsection{Graph construction from Hematoxylin and eosin stain (H\&E) stained images}
We have used a UNet \cite{ronneberger2015unet} based model for detecting the nuclei. Edge features are based on the inter-nucleus distance. We measure the distance between two nuclei as $$dist(i,j) = \sqrt{(x_i-x_j)^2 +(y_i-y_j)^2}$$, where $(x_{i}, y_{i})$ are the co-ordinates of nucleus $n_{i}$.
We form an edge between two nodes i and j, $A_{i,j}$ if their inter-nuclei distance is less than 100 pixels and assign the following weight to the resultant edge in the adjacency matrix (A):

\begin{equation}
    A_{i,j} = 1-\frac{dist(i,j)}{100}
\end{equation}
\begin{figure*}
\centering     
\subfigure[Whole slide tissue image]{\label{fig:a}\includegraphics[width=0.3\linewidth]{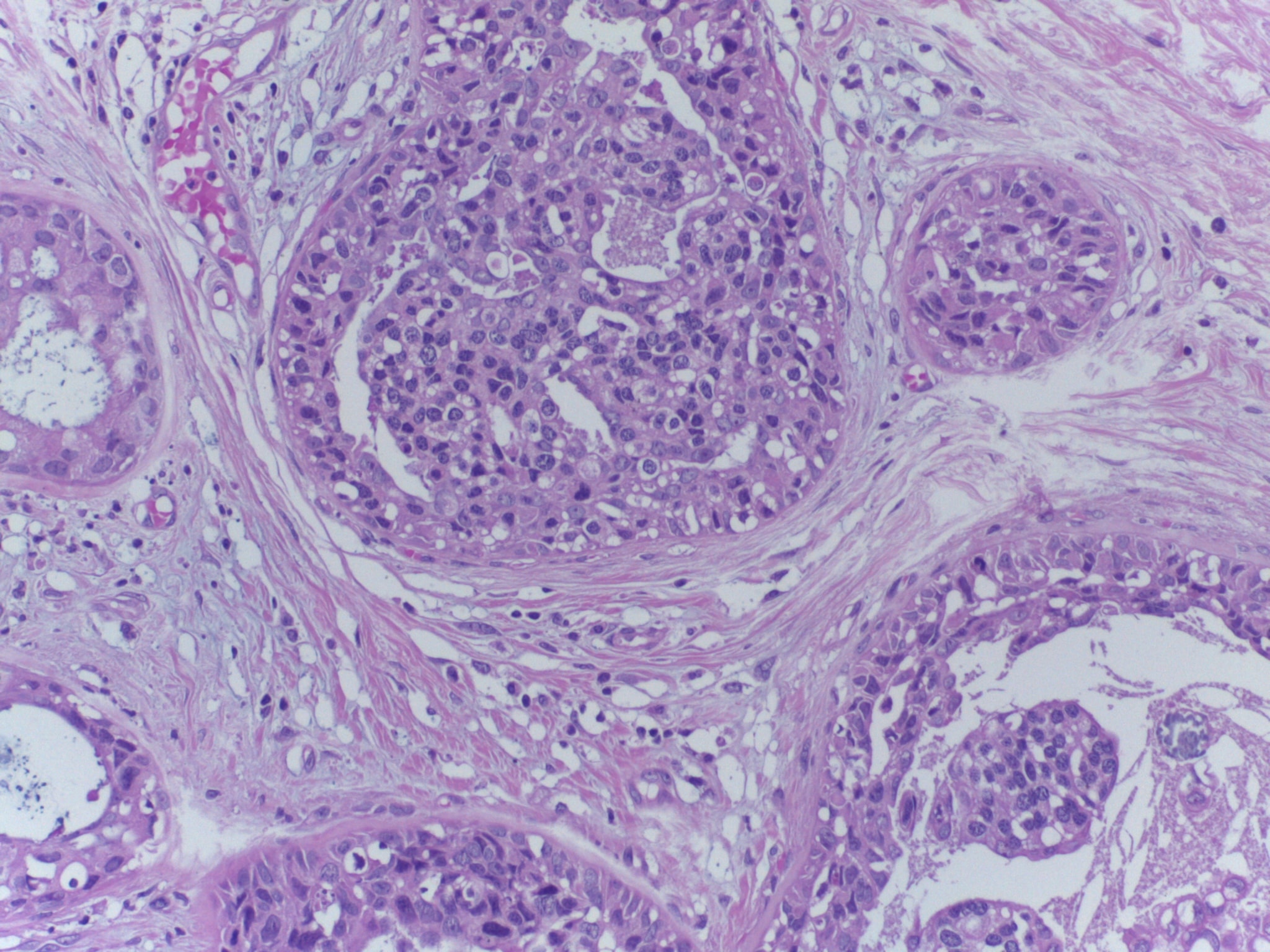}}
\subfigure[Detected nucleus mask]{\label{fig:b}\includegraphics[width=0.3\linewidth]{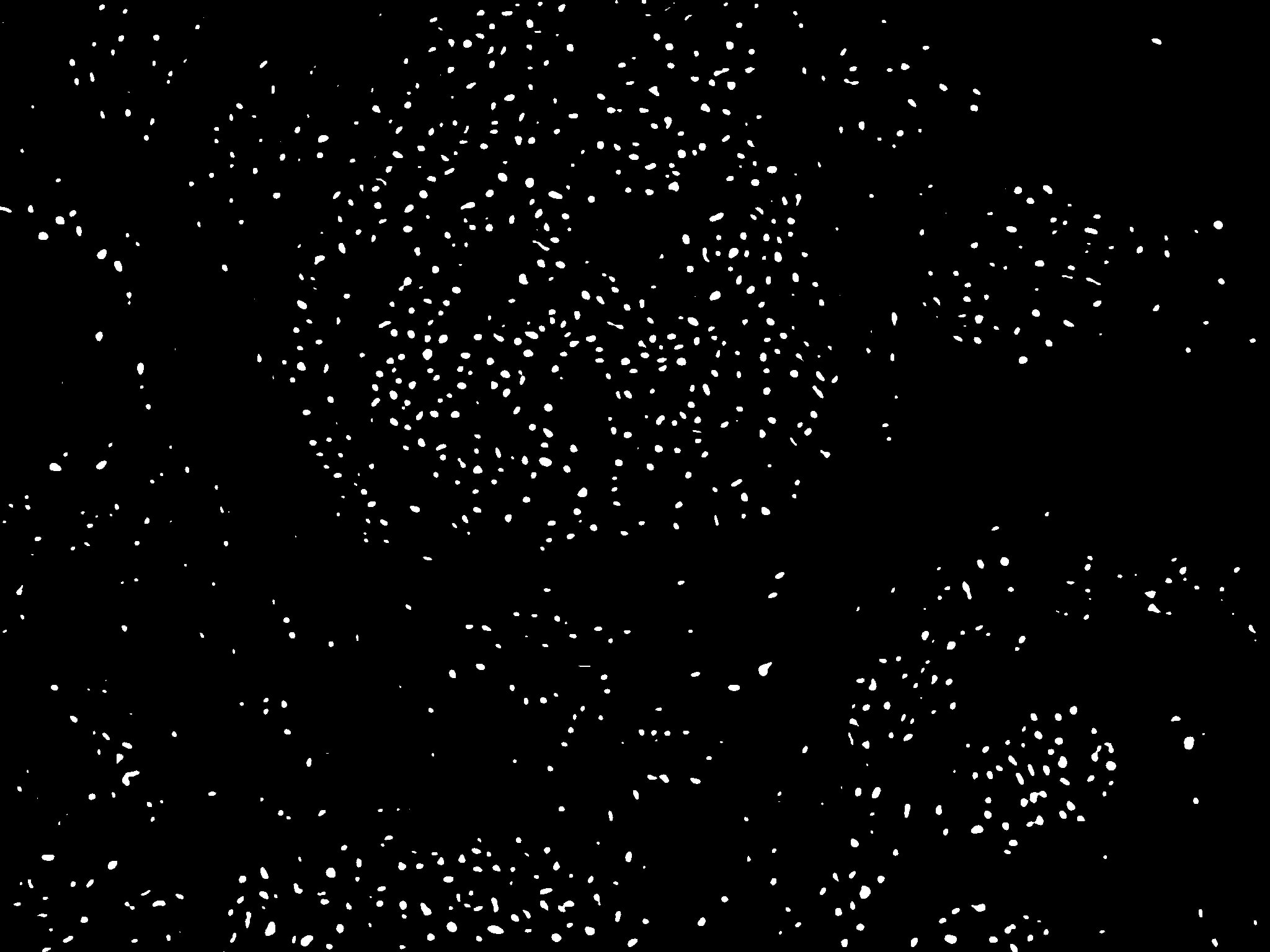}}
\subfigure[Generated graph]{\label{fig:c}\includegraphics[width=0.3\linewidth]{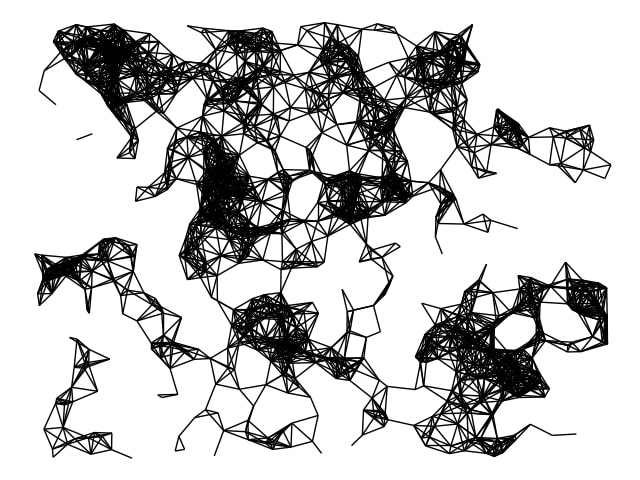}}
\caption{Graph formation: We start with a histopathology image, detect all nuclei using a U-Net, and construct a graph by linking pairs of nuclei closer than a distance threshold.}
\label{combination}
\end{figure*}

\subsection{Robust spatial filtering (RSF)}
Our GCN was adapted from robust spatial filtering (RSF) \cite{such2017robust}. For a graph $G(V, E)$, $V$ is the set of vertices and $E$ is the set of edges and $N$ is the number of nodes. Each vertex and edge can have multiple features.The numbers of features for a vertex and an edge are $F$ and $L$ respectively. The above arrangement allows the set $V$ and $E$ to be represented as tensors such as $V \in \mathcal{R}^{ N\times F}$ and $E \in \mathcal{R}^{N\times N \times L}$ respectively. In RSF, the convolution operation on graphs is given by the following equation:
\begin{equation} \label{eq1}
    \begin{split}
        V_{conv}  &= \sum_{i=1}^{F} H^i V_{in}^i + b, \\
        \text{where,  } H^i &= h_0^i + \sum_{j=1}^{L} h_j^i A_j
    \end{split}
\end{equation}
\\
where, $h_i^j$ and b are learnable parameters and $A_j$ represents the ${j^{th}}$ edge feature of adjacency matrix.  Multiple such filters are used to learn $F^{'}$ vertex features. In RSF, the graph adjacency matrix is not transformed into the spectral domain. Hence the computationally heavy operation of inversion of the Laplacian matrix is avoided. 

For pooling operation, $V_{emb}^{'} \in \mathcal{R}^{N \times N'}$ is derived from the input graph with $V_{in} \in \mathcal{R}^{N \times F}$ and $A \in \mathcal{R} ^{N\times N \times L}$. This operation is similar to convolution operation given in Equation \ref{eq1}. Further, $V_{out} \in \mathcal{R}^{N' \times F}$ and $A_{out} \in \mathcal{R}^{N' \times N' \times F}$ with $N' < N$ is obtained by,
\begin{equation}
    \begin{split}
         V_{emb} &= Softmax(V) \\
         V_{out} &= V_{emb}^T V_{in} \\
         A_{out} &= V_{emb}^TA_{in}V_{emb}
    \end{split}
\end{equation}
\subsection{RSF with edge convolutions (RSF+Edge) }
The convolutional layer in RSF convolves vertex features of neighbor vertices to learn enhanced vertex features. This operation does not exploit the edge features directly. Gadiya et al. \cite{gadiya2018some} proposed a method to learn enhanced vertex as well as edge features. Edge convolutional is performed as per the following equation:
\begin{equation}
    A_{out} = \phi ( W X )
\end{equation}
where $W$ is tensor of learnable parameters and $X$ is obtained by concatenating edge and vertex features of a node and $\phi$ is a monotonic nonlinear activation function.
\subsection{Robust Spatial Filtering with Attention (RSF+Attention)}
We conjectured that an attention mechanism could help rank the graph vertices in their relative order of importance. Attention mechanism is used in neural networks extensively for natural language processing and to a lesser extent for computer vision tasks \cite{xu2015show,ilse2018attention}. In our work, the attention layer was included before the first pooling operation at the input to highlight important nuclei directly, as shown in Figure 1.
\subsection{Visualization}
For the proposed model (RSF+Attention), we used the attention scores for visualization of the importance of individual nuclei. For the models that lacked an attention mechanism, given a trained model $\mathcal{M}$ and a graph $G$, we rank all the nodes based on the drop in classification probability in a manner similar to \cite{zeiler2014visualizing}.  To get a more discernible drop in accuracy, for every node all the 1-hop neighbors along with their edges were also occluded. Occlusion of a node $n_{i}$ creates  a new graph $G_{n_{i}}$ . Classification probability is computed for the occluded graph. The relative drop in probability for the nodes $n_{i}$ gives a measure $score_{i}$ for the importance of each node. We also tested 2-hop and 3-hop occlusion but the results were similar to those of 1-hop. Formally, $score_{i}$ for node $n_{i}$ can be given as,

\begin{equation}
    score_{i}  =  p(\mathcal{M}(G)) - p(\mathcal{M}(G_{n_{i}})) 
\end{equation}

\section{Experiments and Results}
In this section, we show graphs formed from histology images, classification accuracy of using various GCN architectures, and visualization of highlighted nuclei.

\begin{table*}
\begin{tabular}{llll}
\hspace{2cm}
Original image \hspace{1.5cm} & Detected nucleus map \hspace{1.7cm} & RSF+edge \hspace{1.8cm} & RSF+attention 
\end{tabular}
\end{table*}
\ExplSyntaxOn
\NewDocumentEnvironment{places}{mm}
 {
  \setlength{\tabcolsep}{2pt} 
  \dim_set:Nn \l_places_width_dim
  {
    (#1-\ht\strutbox-\dp\strutbox-2pt)/(#2)
  }
  \begin{tabular}{r @{\hspace{2pt}} *{#2}{c}}
 }
 {
  \end{tabular}
 }
\NewDocumentCommand{\place}{mm}
 {
  \seq_set_from_clist:Nn \l_places_images_in_seq { #2 }
  \seq_set_map:NNn \l_places_images_out_seq \l_places_images_in_seq { \places_set_image:n {##1} }
  \seq_put_left:Nn \l_places_images_out_seq
  {
    \begin{tabular}{c}\rotatebox[origin=c]{90}{\strut#1}\end{tabular}
  }
  \seq_use:Nn \l_places_images_out_seq { & } \\ \addlinespace
 }
\dim_new:N \l_places_width_dim
\seq_new:N \l_places_images_in_seq
\seq_new:N \l_places_images_out_seq
\cs_new_protected:Nn \places_set_image:n
 {\makebox[\l_places_width_dim]
  { \begin{tabular}{c}
    \includegraphics[
      width=\l_places_width_dim,
      height=\l_places_width_dim,
      keepaspectratio,
    ]{#1}
    \end{tabular}
  }
 }
\ExplSyntaxOff
\begin{figure*}
\centering
\begin{places}{0.9\textwidth}{4}
\place{In-situ }{
  image_nucleus_graph/is10.jpg,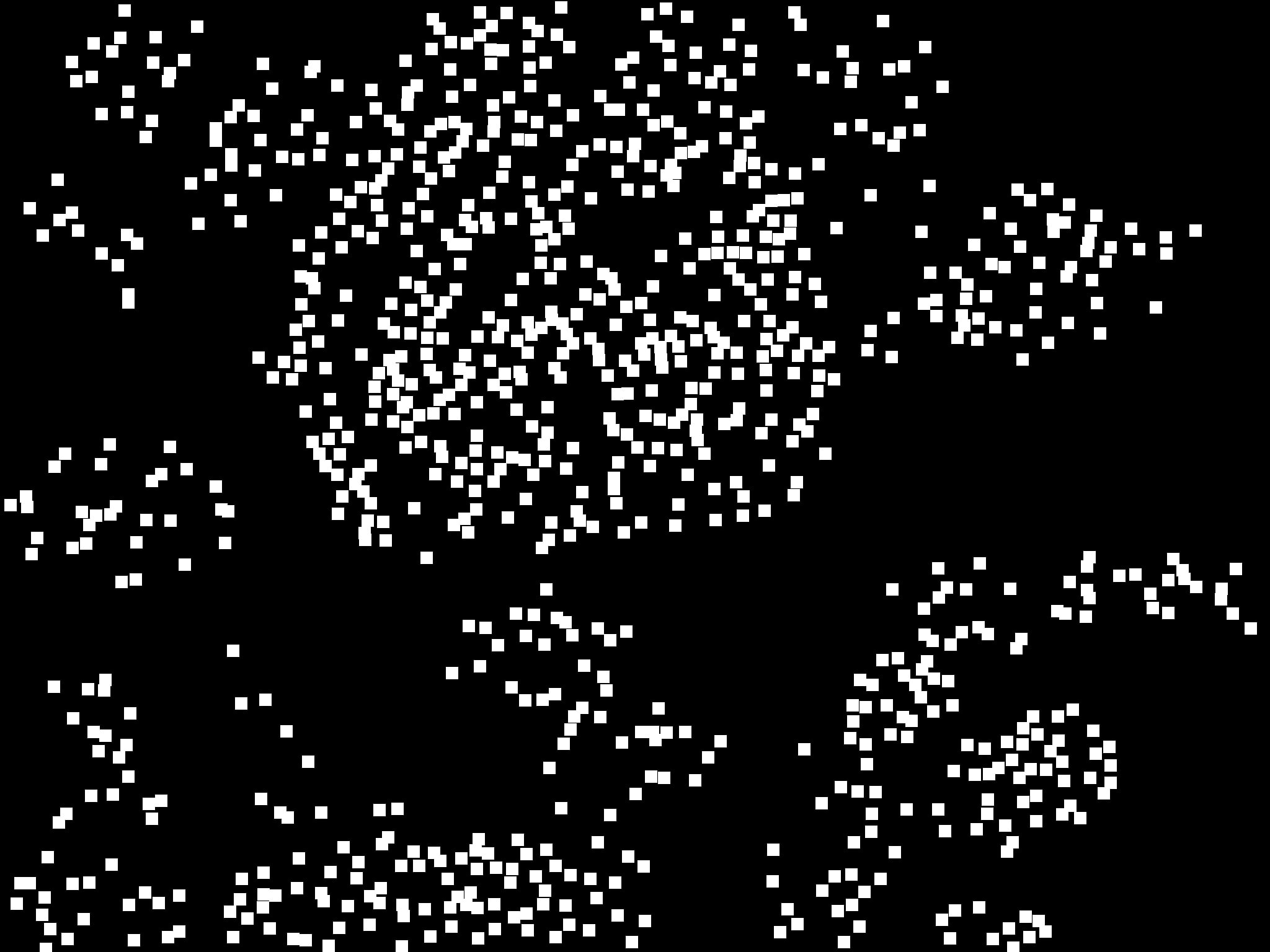,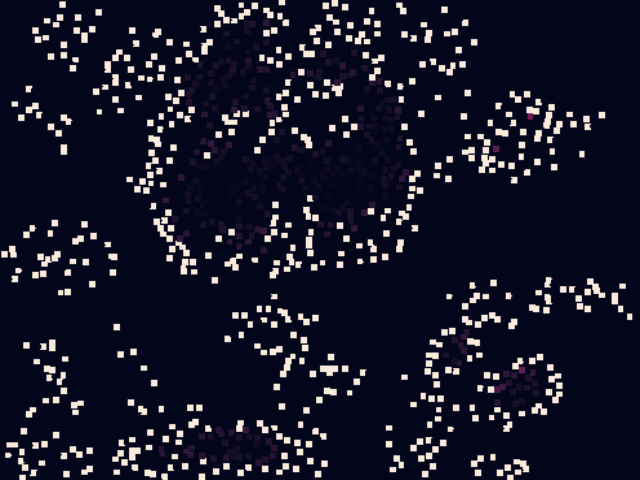,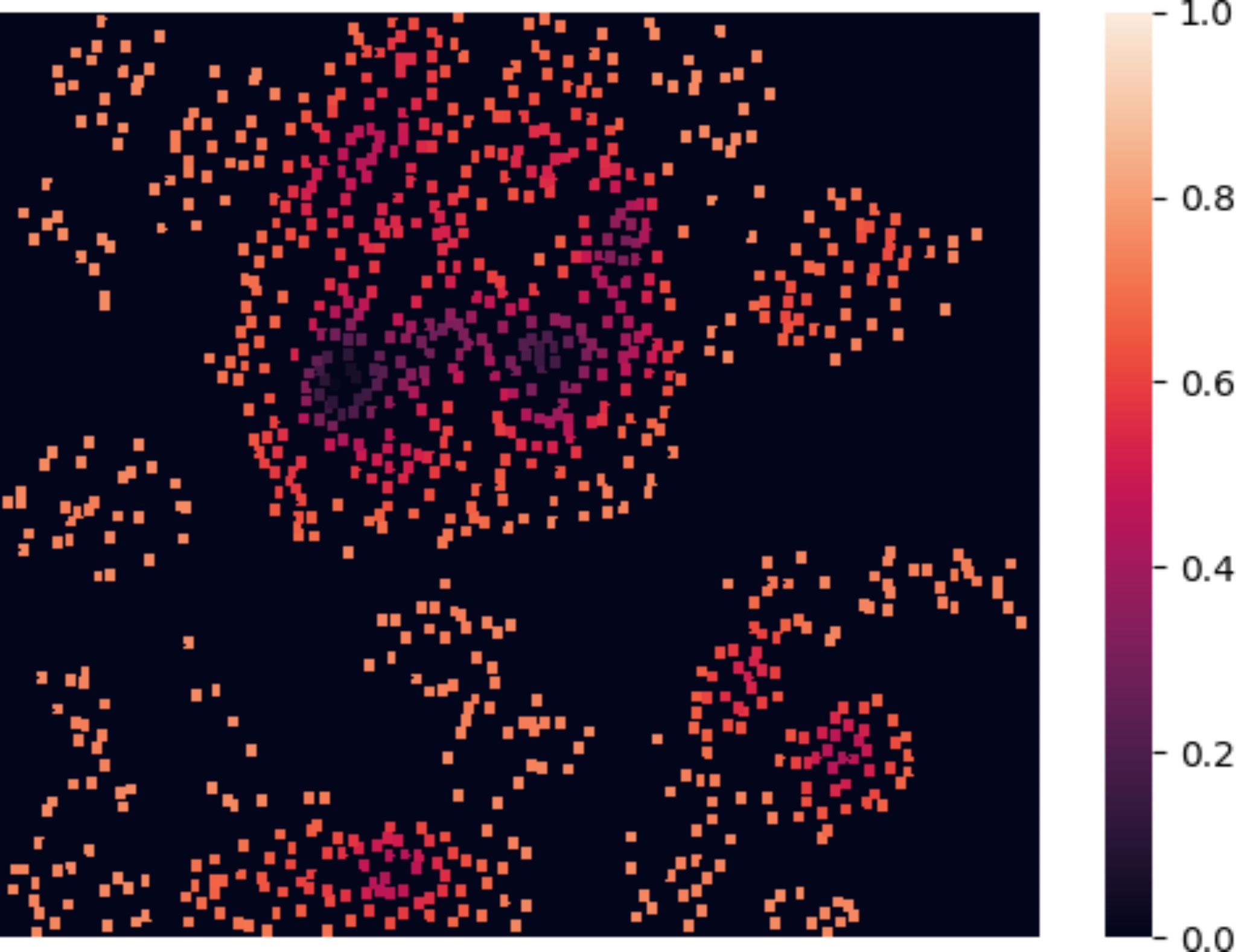
}
\place{Invasive}{
  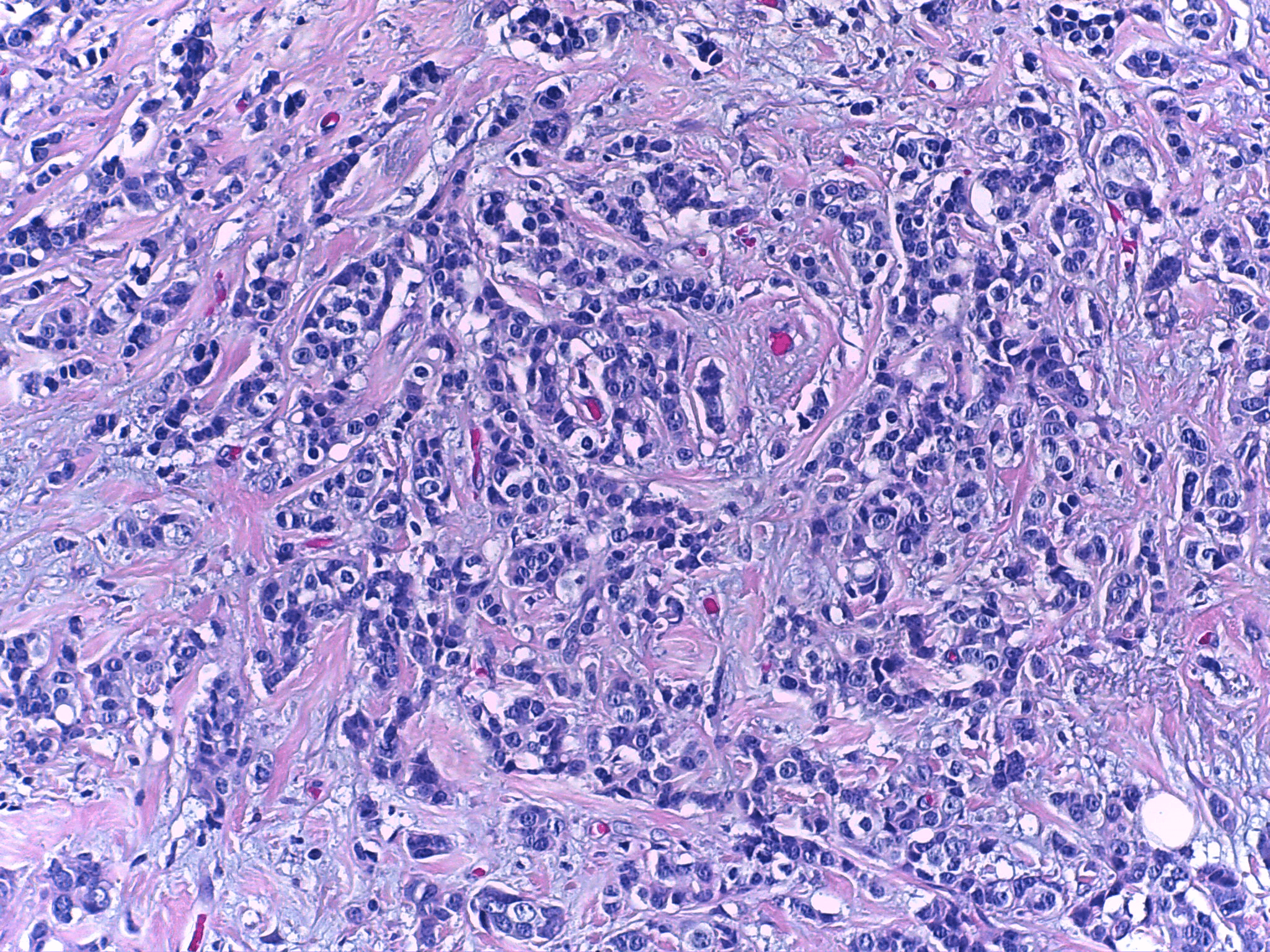,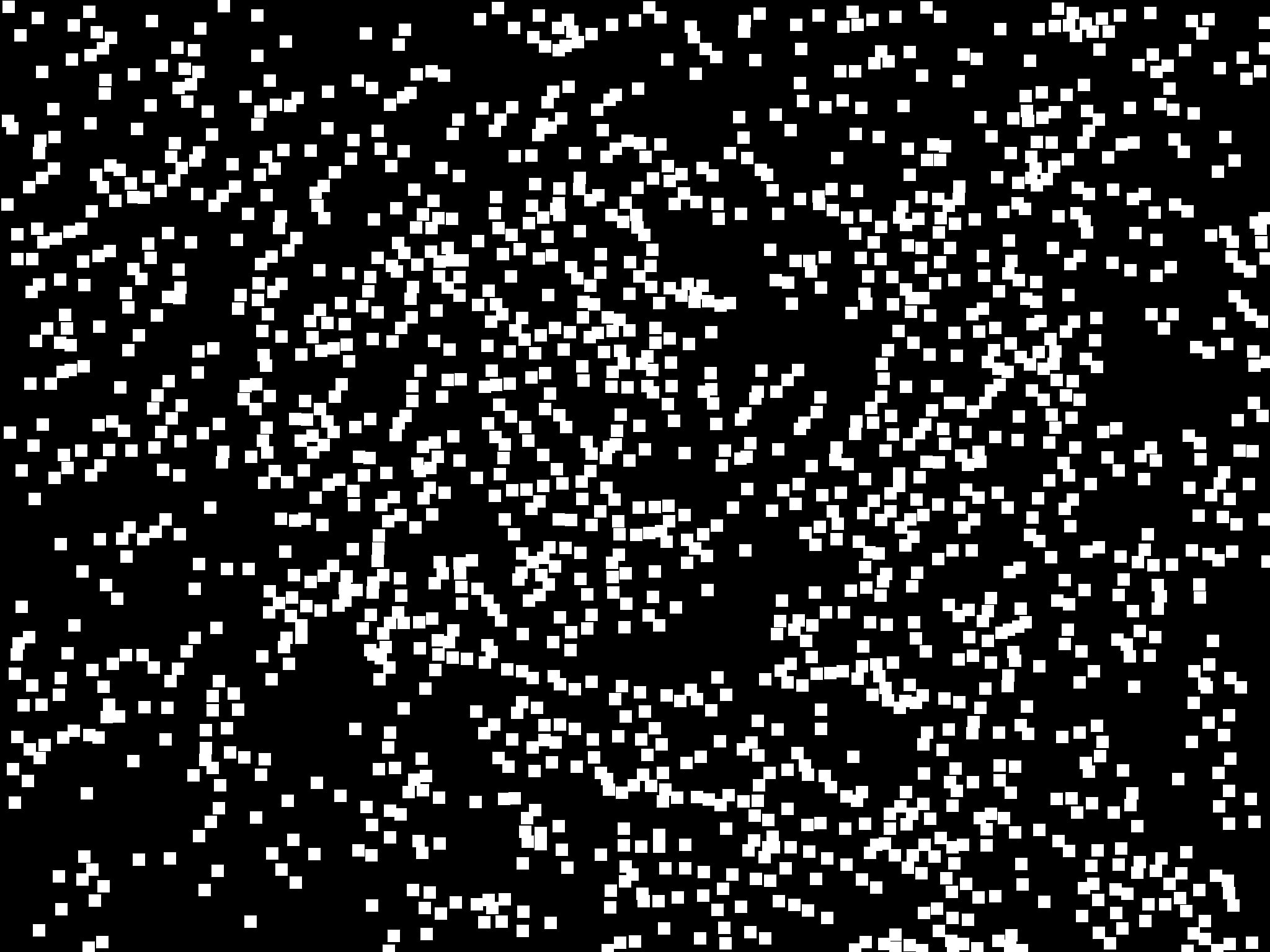,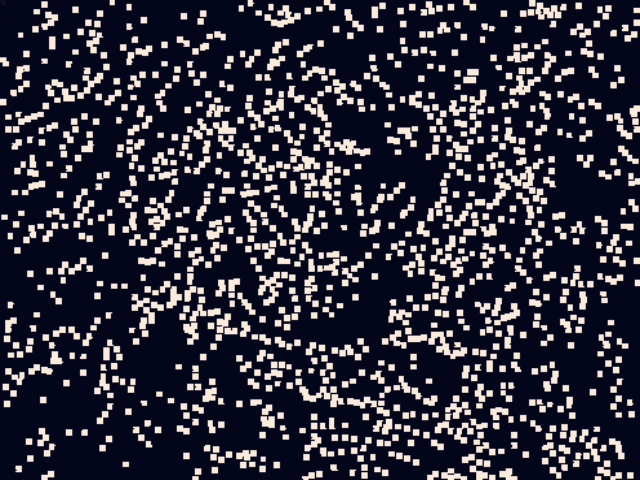,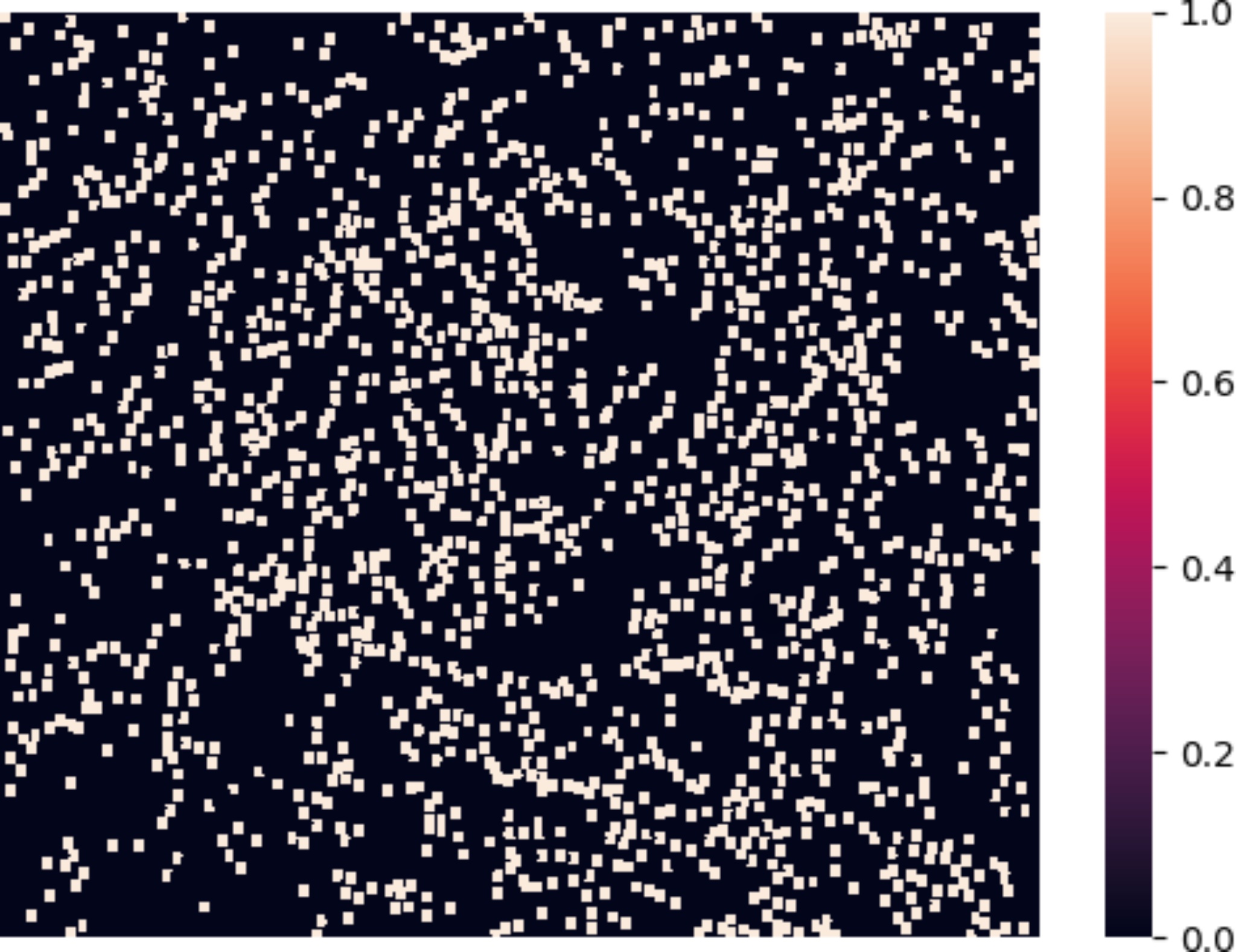
}
\place{Gleason 3}{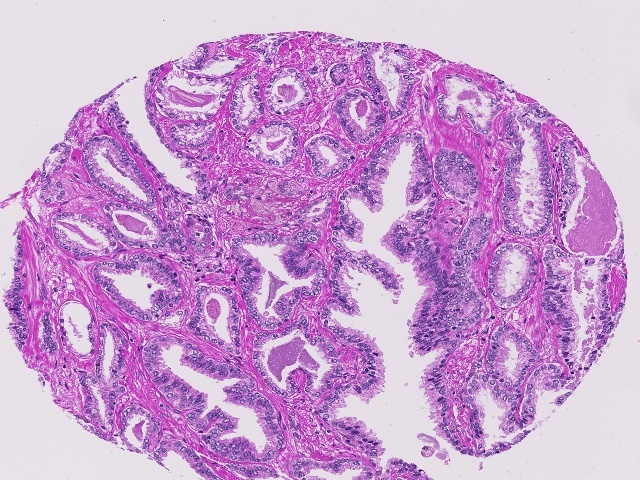,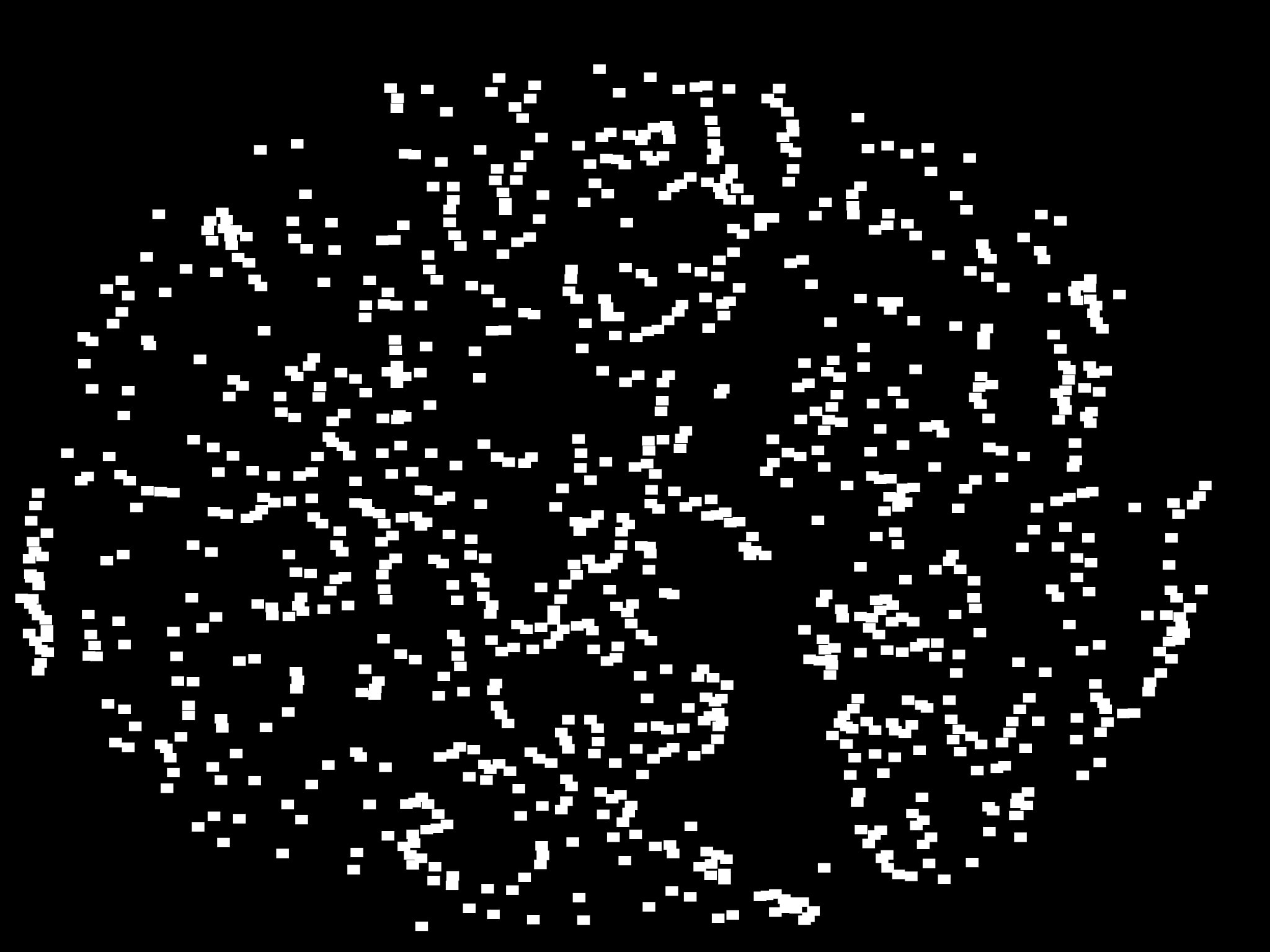,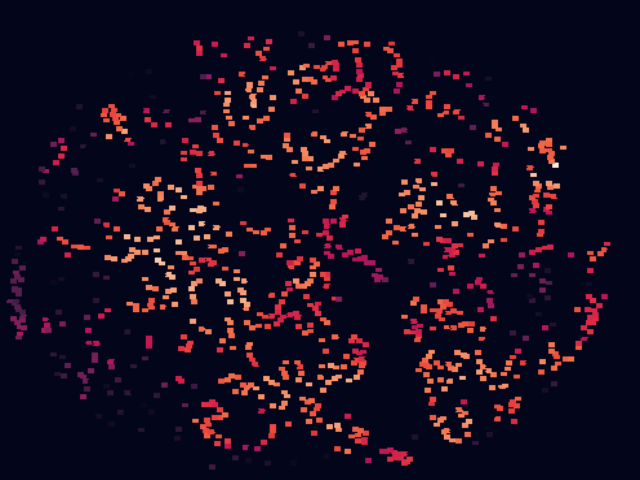,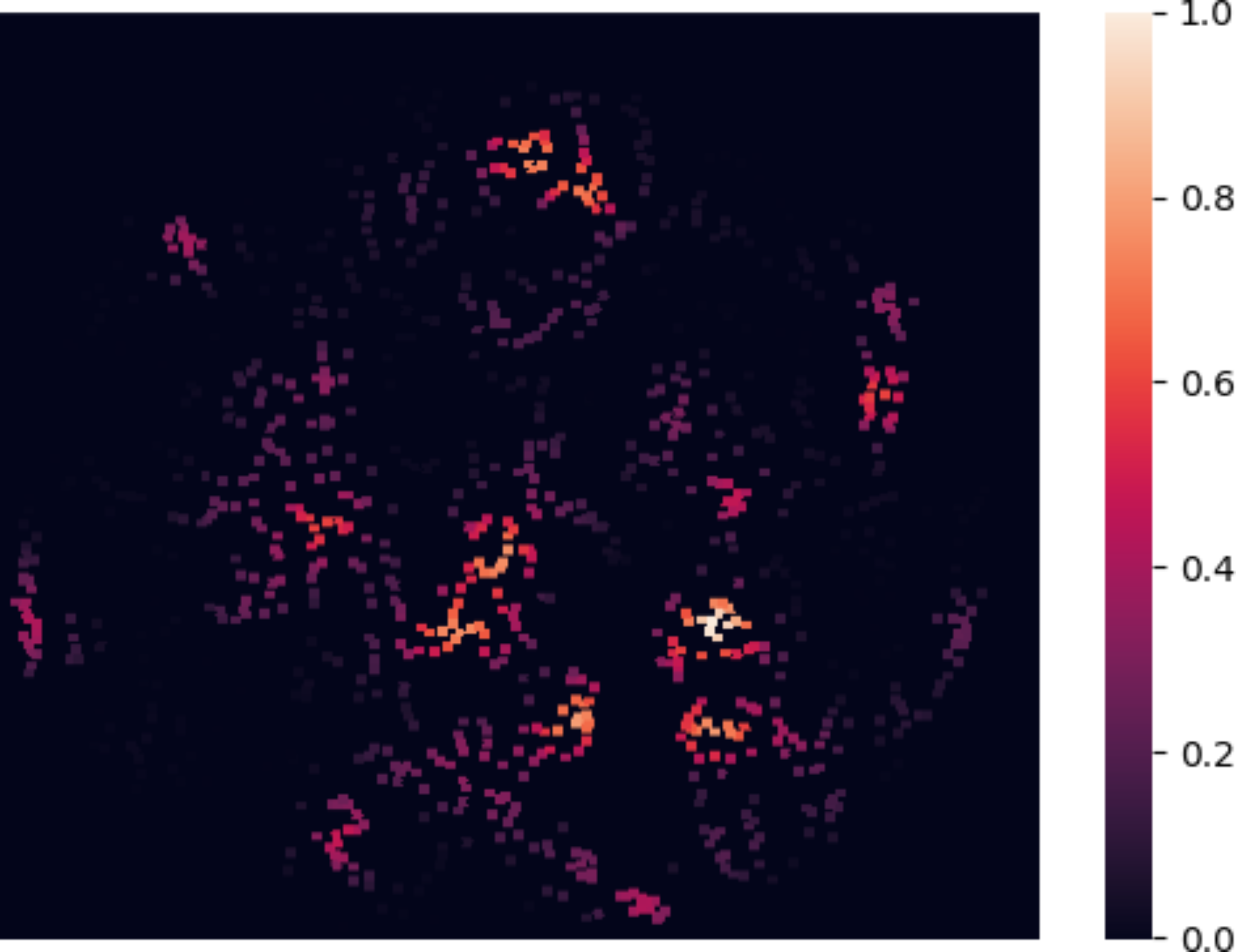}
\place{Gleason 4}{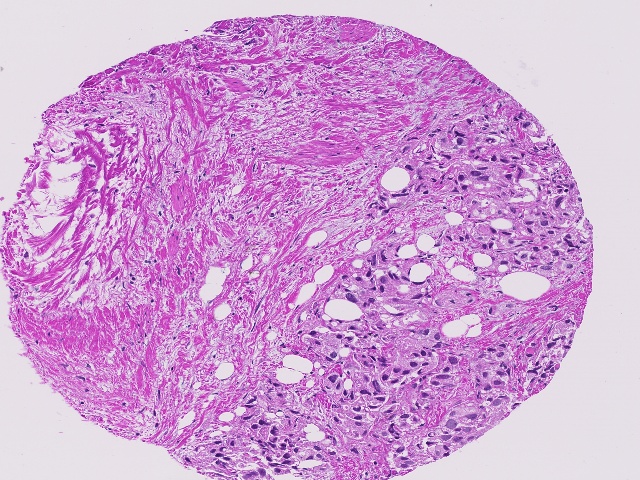,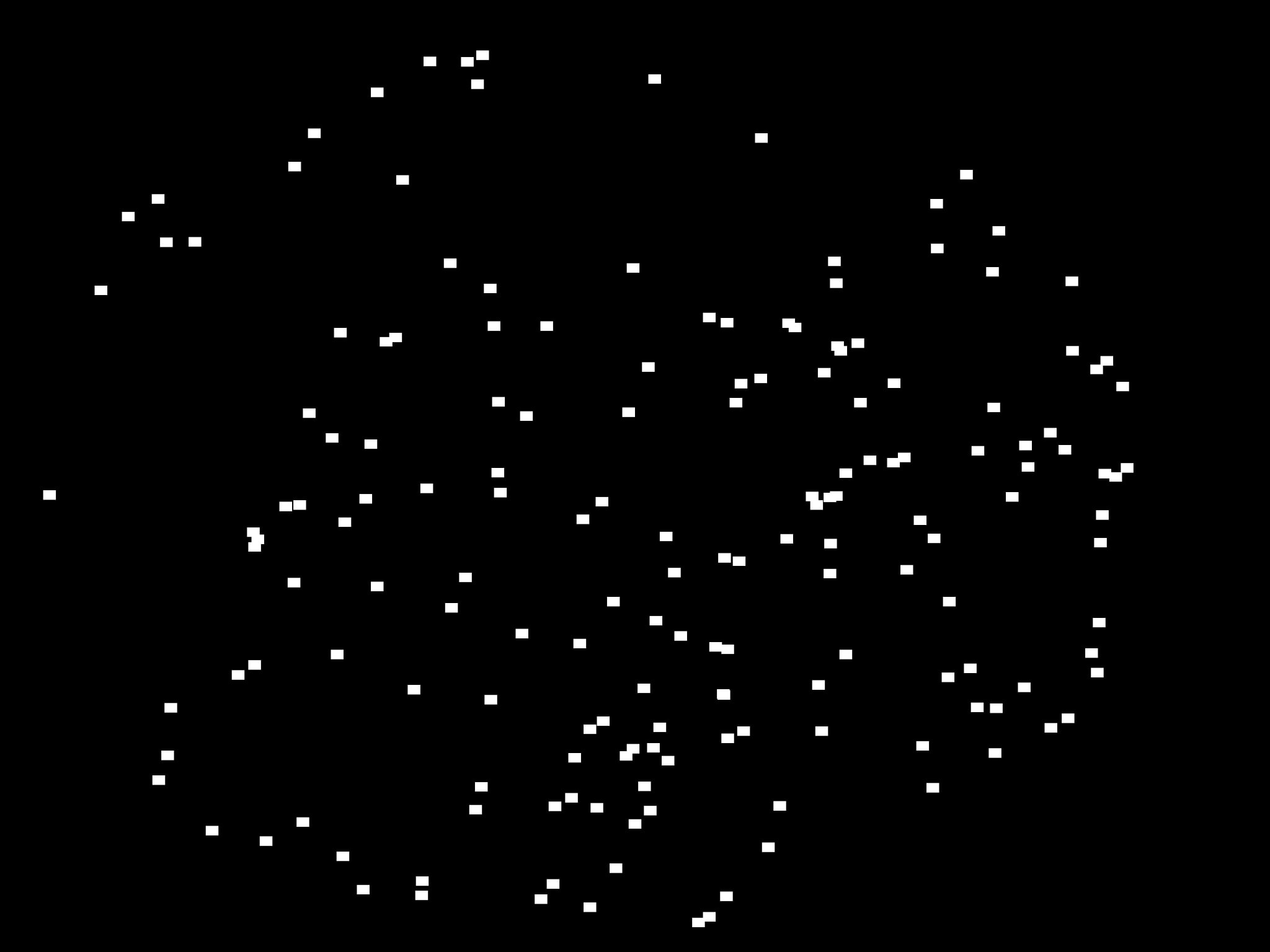,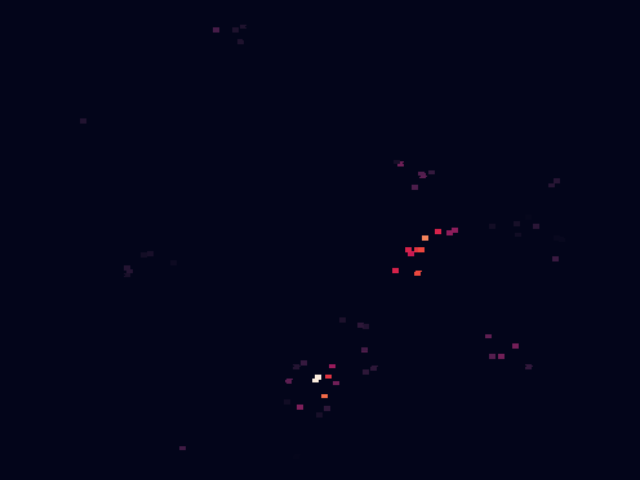,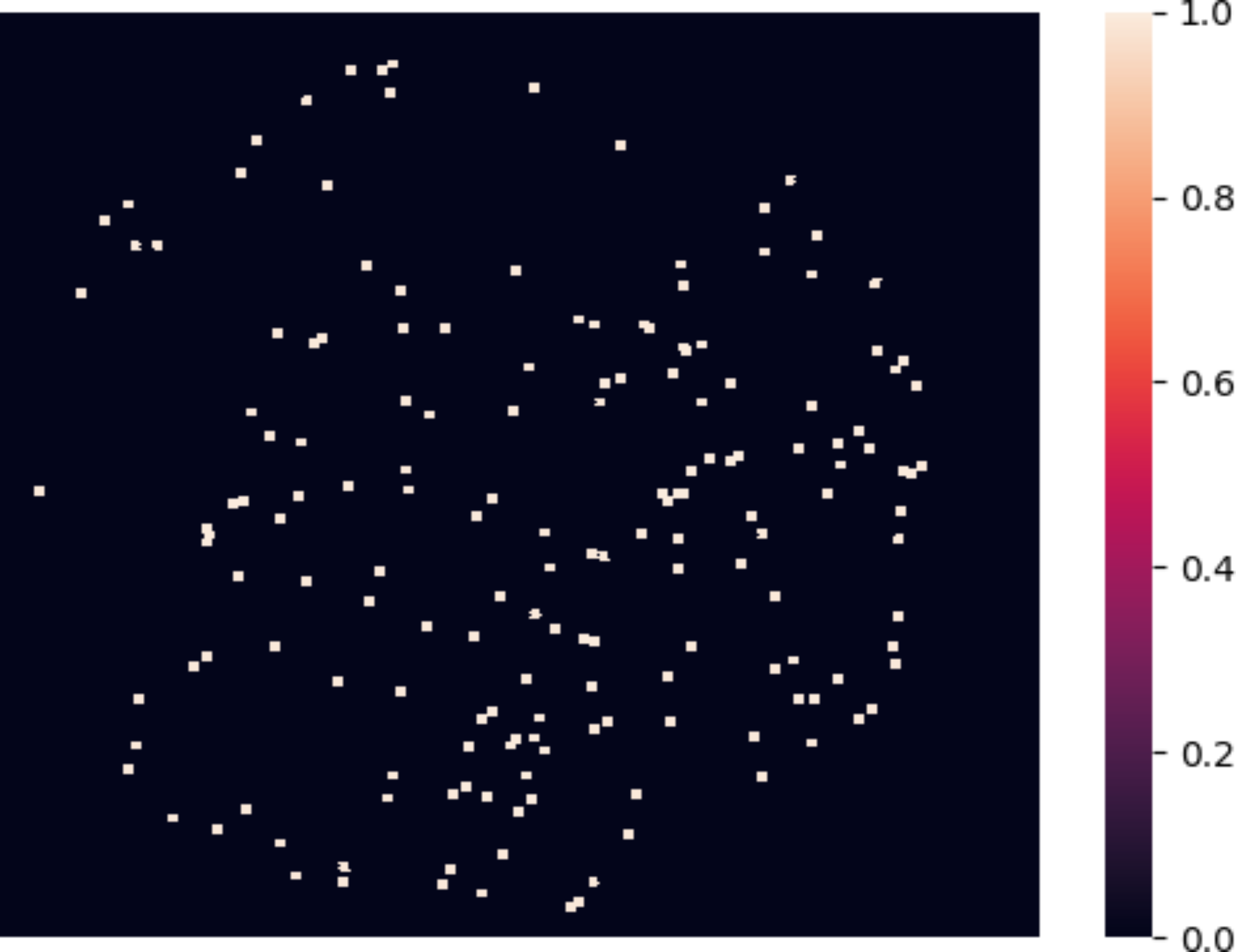}
\end{places}
\caption{Comparison of visualization of RSF+edge and the proposed RSF+attention: Using RSF+attention, while nuclei on gland boundary are relatively more highlighted in the in-situ breast cancer (first row), all cancerous nuclei are highlighted in invasive breast cancer (first row). Similarly, the gland shapes are prominently highlighted in Gleason 3 prostate cancer (third row) as opposed to all cancer cells being highlighted in Gleason 4 prostate cancer (bottom row).  The scale on the right shows color scale for the relative importance of nuclei.}
\label{Insitu(1)}
\end{figure*}

\subsection{Graphs from H\&E stained histopathology images}  
Each image produces a graph with a different number of nodes. For BACH and prostate cancer Gleason grade datasets, the average number of nodes in a graph was 1546 and 613, respectively. Figure \ref{combination} shows an example of transforming H\&E stained histopathology image to a graph.  

\subsection{Classification of breast and prostate cancers}
\begin{table}
\centering
\begin{tabular}{|c|c|c|}
\hline
\textbf{RSF}           & \textbf{RSF + Edge}    & \textbf{RSF + Attention} \\ \hline
Vertex Conv 1 & Vertex Conv 1 & Vertex Conv 1             \\
Vertex Conv 2 & Vertex Conv 2 & Vertex Conv 2 + Attn \\ \hline
Pooling 1     & Pooling 1     & Pooling 1                 \\ \hline
              & Edge Conv 1   &                           \\
Vertex Conv 3 & Vertex Conv 3 & Vertex Conv 3             \\
              & Edge Conv 2   &                           \\ \hline
Pooling 2     & Pooling 2     & Pooling 2                 \\ \hline
              & Edge Conv 3   &                           \\
FC - 1        & FC - 1        & FC - 1                    \\
FC - 2        & FC - 2        & FC - 2                    \\
FC - 3        & FC - 3        & FC - 3                    \\ \hline
\end{tabular}
\caption{The architecture of techniques implemented}
\label{architecture}
\end{table}
\begin{table}
\centering
\begin{tabular}{
>{\columncolor[HTML]{FFFFFF}}c |
>{\columncolor[HTML]{FFFFFF}}c |
>{\columncolor[HTML]{FFFFFF}}c }
\hline
\textbf{Model} & \textbf{BACH} & \textbf{Gleason} \\ \hline
RSF Based & 94\% & 97\% \\
RSF + Edge & 92\% & 97\% \\
RSF + Attention & 90\% & 97\%
\end{tabular}
\caption{Accuracy results for different techniques}
\label{results}
\end{table}
We trained the three models described in the previous section, viz. robust spatial filtering (RSF), robust spatial filtering with edge convolution (RSF+Edge), and robust spatial filtering with attention (RSF+Attention). All models were trained for approximately 50 epochs with a learning rate of 0.01 using the Adam optimizer. The architectures of the three models are given in table \ref{architecture}. Table \ref{results} shows that classification accuracy for the three models was quite comparable to each other. All the models contained nearly 300,000 parameters.

\subsection{Visualization} 
We now present the visualization produced by occlusion and attention mechanisms. We performed occlusion experiments on predictions of RSF and RSF+Edge models on the breast and prostate cancer datasets. Visualization produced by these models were nearly the same, so we have omitted the results from the former due to space constraints. The images in the first row correspond to in-situ subtype in breast cancer from BACH dataset. We can see that nuclei on the outer layer of the gland are highlighted by the occlusion experiments. Also, in the second row, which corresponds to the invasive class in BACH dataset, nearly all the nuclei are highlighted. Outer linings are crucial for in-situ classification and where as for invasive cancer is spread across the entire region. These are the characteristics of in-situ and invasive histologies that are correctly captured by the occlusion and attention experiments.   In the last two rows, visualization results for the prostate cancer Gleason grade dataset are shown. In these images, nuclei of the glands that lose their structure are highlighted, as we expected them to be. The images in the last column of Figure \ref{Insitu(1)} are visualization results from RSF+Attention model. These results were verified by expert pathologists and visibly better at highlighting the above mentioned features.

\section{Conclusion}
We occluded nuclei clusters and exploited an attention layer in a graph convolutional neural network to highlight nuclei in histopathology slides and visualized the results on a breast cancer and a prostate cancer datasets. The proposed methods provide a notably more interpretable map depicting the contribution of each nucleus and its neighborhood in the final diagnosis. The presented results provide a way to explain the new patterns the deep learning models found on the tissue images. The proposed techniques not only open a path for the verification of the existing practices in pathology but suggest a way to generate new knowledge on where to focus to find meaningful differences between tissue classes, for example, those that may have different disease or treatment outcome.

\section*{Acknowledgment}
Authors would like to thank Nvidia Corporation for donation of GPUs used for this research.

\bibliography{main}
\bibliographystyle{IEEEtran}
\end{document}